\documentclass[%
 reprint,
superscriptaddress,
 amsmath,amssymb,
 aps,
]{revtex4-2}

\usepackage{graphicx}
\graphicspath{ {Figures/} }
\usepackage{dcolumn}
\usepackage{bm}
\usepackage[hyperindex,breaklinks]{hyperref}
\hypersetup{colorlinks=true}
\usepackage{breakurl}
\usepackage[mathlines]{lineno}
\usepackage[caption=false]{subfig}
\usepackage{booktabs}

\newcommand{\nuebar}{$\overline{\nu}_{e}$ }
\newcommand{\nuebars}{$\overline{\nu}_{e}$'s }
\newcommand{\alphan}{($\alpha, n$) }
\newcommand{\dmResult}{($7.96^{+0.48}_{-0.42}$) $\times$ 10$^{-5}$ }
\newcommand{\sstResRight}{$0.62^{+0.16}_{-0.40}$}
\newcommand{\geoResult}{$73^{+47}_{-43}$ }

\begin{document}

\preprint{APS/123-QED}

\title{Measurement of reactor antineutrino oscillation at SNO+}

\author{ M.\,Abreu }
\affiliation{\it Laborat\'{o}rio de Instrumenta\c{c}\~{a}o e  F\'{\i}sica Experimental de Part\'{\i}culas (LIP), Av. Prof. Gama Pinto, 2, 1649-003, Lisboa, Portugal }
\affiliation{\it Universidade de Lisboa, Instituto Superior T\'{e}cnico (IST), Departamento de F\'{\i}sica, Av. Rovisco Pais, 1049-001 Lisboa, Portugal }
\author{ V.\,Albanese }
\affiliation{\it Queen's University, Department of Physics, Engineering Physics \& Astronomy, Kingston, ON K7L 3N6, Canada }
\affiliation{\it Laurentian University, School of Natural Sciences, 935 Ramsey Lake Road, Sudbury, ON P3E 2C6, Canada }
\author{ A.\,Allega }
\affiliation{\it Queen's University, Department of Physics, Engineering Physics \& Astronomy, Kingston, ON K7L 3N6, Canada }
\author{ R.\,Alves }
\affiliation{\it Laborat\'{o}rio de Instrumenta\c{c}\~{a}o e F\'{\i}sica Experimental de Part\'{\i}culas (LIP), 3004-516, Coimbra, Portugal }
\author{ M.\,R.\,Anderson }
\affiliation{\it Queen's University, Department of Physics, Engineering Physics \& Astronomy, Kingston, ON K7L 3N6, Canada }
\author{ S.\,Andringa }
\affiliation{\it Laborat\'{o}rio de Instrumenta\c{c}\~{a}o e  F\'{\i}sica Experimental de Part\'{\i}culas (LIP), Av. Prof. Gama Pinto, 2, 1649-003, Lisboa, Portugal }
\author{ L.\,Anselmo }
\affiliation{\it SNOLAB, Creighton Mine \#9, 1039 Regional Road 24, Sudbury, ON P3Y 1N2, Canada }
\author{ J.\,Antunes }
\affiliation{\it Laborat\'{o}rio de Instrumenta\c{c}\~{a}o e  F\'{\i}sica Experimental de Part\'{\i}culas (LIP), Av. Prof. Gama Pinto, 2, 1649-003, Lisboa, Portugal }
\affiliation{\it Universidade de Lisboa, Instituto Superior T\'{e}cnico (IST), Departamento de F\'{\i}sica, Av. Rovisco Pais, 1049-001 Lisboa, Portugal }
\author{ E.\,Arushanova }
\affiliation{\it Queen Mary, University of London, School of Physics and Astronomy,  327 Mile End Road, London, E1 4NS, UK }
\author{ S.\,Asahi }
\affiliation{\it Queen's University, Department of Physics, Engineering Physics \& Astronomy, Kingston, ON K7L 3N6, Canada }
% ...
\collaboration{ The SNO+ Collaboration }


\begin{abstract}
The SNO+ collaboration reports its second spectral analysis of reactor antineutrino oscillation using 286 tonne-years of new data.  The measured energies of reactor antineutrino candidates were fitted to obtain the second-most precise determination of the neutrino mass-squared difference $\Delta m^2_{21}$ = \dmResult eV$^2$. 
Constraining $\Delta m^2_{21}$ and $\sin^2\theta_{12}$ with measurements from long-baseline reactor antineutrino and solar neutrino experiments yields $\Delta m^2_{21}$ = ($7.58^{+0.18}_{-0.17}$) $\times$ 10$^{-5}$~eV$^2$ and $\sin^2\theta_{12} = 0.308 \pm 0.013$. 
This fit also yields a first measurement of the flux of geoneutrinos in the Western Hemisphere, with \geoResult TNU at SNO+. 
\end{abstract}

\maketitle


\textit{Introduction}.  
Neutrino oscillation is well established through measurements of neutrino rates and energy spectra from particle accelerators, nuclear reactors, the atmosphere, and the Sun.  These measurements provide consistent values for the three neutrino oscillation angles $\theta_{ij}$ and two mass-squared differences $\Delta m^2_{ij} \equiv m^2_i - m^2_j$, where $i$ and $j$ are 1, 2, or 3 ($i \neq j$).  
Leading measurements of $\Delta m^2_{21}$, $\theta_{13}$, and $\Delta m^2_{32}$ have been obtained with reactor antineutrinos~\cite{kamland_on_off,daya_bay_theta13_dm232,reno_theta13_dm232,double_chooz_theta13_dm232}.
The KamLAND experiment has made the most precise measurement of $\Delta m^2_{21}$, along with a less sensitive measurement of $\sin^2\theta_{12}$~\cite{kamland_on_off}. In contrast, solar neutrino experiments have provided the most precise measurement of $\sin^2\theta_{12}$ while being less sensitive to $\Delta m^2_{21}$. The KamLAND measurement of $\Delta m^2_{21}$ is currently in 1.5$\sigma$ tension with the result of a combined analysis of all available solar neutrino data performed by Super-K~\cite{Super-Kamiokande:2023jbt}. 

SNO+ has previously published results studying reactor antineutrinos, including the first evidence of reactor antineutrino detection with a water Cherenkov detector~\cite{snoplus_water_antinu} and an initial measurement of reactor antineutrino oscillation with the detector nearly half-filled with liquid scintillator~\cite{snoplus_partial_antinu}.  
Radioactive decays within the Earth also produce antineutrinos, which have been observed by the KamLAND and Borexino liquid scintillator detectors in Japan and Italy, respectively~\cite{kamland_geonu, borexino_geonu}. 
This Letter presents a measurement of $\Delta m^2_{21}$ and $\sin^2\theta_{12}$ using reactor antineutrinos, and initial constraints on the geoneutrino flux, based on the first data collected by SNO+ as a fully-filled liquid scintillator detector. 

\textit{Data}.  
The SNO+ detector now contains about 780~tonnes of liquid scintillator within its 6.0-m radius spherical acrylic vessel (AV). Light produced by interactions in the scintillator is detected by 9362 inward-facing photomultiplier tubes (PMTs) at a radius of about 8.5~m. The scintillator volume is shielded by ultrapure water between the AV and PMTs and also beyond, where outward-looking PMTs are used to detect muons. 
The SNO+ detector is described in detail in Ref.~\cite{snoplus_detector}. 

The livetime of the dataset used in this analysis is 134.4~days, collected between May 17, 2022, and March 14, 2023. During this period, the scintillator was linear alkylbenzene (LAB) with 2,5-Diphenyloxazole (PPO) at a concentration of 2.2~g/L. 
The collected scintillation light yield was measured to be about 210~`clean' PMT hits/MeV, which depends on the number of working channels and includes the removal of hits due to electronic noise and exclusion of channels that are not considered well calibrated in terms of charge and timing.  The trigger threshold for this dataset was around 20~PMT hits, corresponding to approximately 0.09~MeV, well below the energy of the reactor antineutrino signals of interest. 

The properties of particle interactions are inferred using the times and locations of hit PMTs. 
The time of flight of the photons detected by the PMTs is used to reconstruct the interaction position.  
The number of hit PMTs is approximately proportional to the energy deposited by a particle. The position-dependent efficiency to detect photons and the probability of detecting multiple photons on individual PMTs are accounted for in the energy reconstruction using Monte Carlo (MC) simulations. 
The resolution of the reconstructed energy $E$ is about 6.5\%/$\sqrt{E}$ for an electron at the center of the detector in this dataset.  The reconstructed position resolution for a 2.5-MeV electron at the center is 12~cm in each of the three Cartesian axes.

\textit{Calibrations}.  
Calibrations were performed using $^{214}$Bi $\beta$ decays and $^{214}$Po $\alpha$ decays identified by their delayed coincidence.  These decays were sourced by the $^{238}$U intrinsic to the detector as well as ingress of $^{222}$Rn into the scintillator. The selection criteria described in Table~\ref{tab:cuts}, which include the time ($\Delta t$) and distance ($\Delta r$) between reconstructed interactions, provide a highly-pure sample of $^{214}$BiPo coincidences. 

\begingroup
\setlength{\tabcolsep}{6pt} 
\renewcommand{\arraystretch}{1} 
\begin{table}[tbp]
    \centering
    \caption{\label{tab:cuts}Selection criteria applied to data and simulations to select $^{214}$BiPo and Reactor-$\overline{\nu}$ IBD coincidence events. See text for parameter definitions.}
\begin{tabular}{l*4c}
\toprule \toprule
 & \multicolumn{2}{c}{$^{214}$BiPo} & \multicolumn{2}{c}{Reactor-$\overline{\nu}$ IBD} \\
\midrule
          & Prompt & Delayed & Prompt & Delayed\\ \midrule
$E$ (MeV) & 1.25-3.0 & 0.7-1.1 & 0.9-9.0 & 1.85-2.5\\
$R$ (m)   & 0-4.0    & 0-4.0   & 0-5.7   & 0-5.7\\
\midrule
$\Delta r$ (m)   & \multicolumn{2}{c}{0-1.0} & \multicolumn{2}{c}{0-2.5} \\
$\Delta t$ ($\mu$s) & \multicolumn{2}{c}{3.7-1000}  & \multicolumn{2}{c}{0-2000}\\
LR & \multicolumn{2}{c}{} & \multicolumn{2}{c}{$>$-3.5}\\
\bottomrule \bottomrule
\end{tabular}
\end{table}
\endgroup

The time profile of the scintillator is modeled for both $\alpha$ and $\beta$ particles as a sum of exponentials with a single rise time parameter. The decay constants and amplitudes are tuned to provide the best match between the time-of-flight-corrected hit times in data and Monte Carlo (MC) simulation. This calibration is performed for $\beta$'s and $\alpha$'s separately since $\alpha$'s produce a broader time profile. 

The scintillator light yield and quenching are described with Birks' law, which is tuned using the PMT hit spectra of the $^{214}$BiPo coincidence events, following the procedure described in the previous analysis~\cite{snoplus_partial_antinu}.  A single light yield is tuned by matching MC simulations of $^{214}$Bi to data, while the Birks' constant for $\beta$'s, $\alpha$'s, and protons are each modeled with a distinct, tuned value. The newly-fit values are consistent with those of the previous analysis. 

After these calibrations were performed, a residual nonuniformity of the energy scale was observed as a function of position nearer the AV. Empirical corrections for both data and simulation were created by fitting the mean energies of the selected $^{214}$BiPo in bins of the vertical position $z$ and the squared horizontal radius $\rho^2 \equiv (x^2 + y^2)$. 
A bilinear interpolation function was used to provide a continuous correction throughout the volume of the AV. After applying the correction, the uncertainties on the energy scale and resolution were evaluated as differences between the data and simulation to be 1.8\% and 4.4\%$\times \sqrt{E}$ (relative), respectively. These uncertainties were consistent between the $\beta$'s and $\alpha$'s.  

\textit{Antineutrino selection}. 
Nuclear reactors produce a large, pure flux of \nuebar with energies up to around 10~MeV in the beta decays of nuclear fission products. These antineutrinos can be detected via the inverse beta decay (IBD) interaction on hydrogen atoms in the detector medium: $\overline{\nu}_{e} + p \rightarrow e^+ + n$. This process has a 1.8-MeV threshold and produces a positron, which quickly annihilates with an electron in the medium, depositing a total energy of $E_{\text{dep}} \approx E_{\overline{\nu}_e} - 0.8 \text{ MeV}$ in the detector. The neutron thermalizes and captures on hydrogen with a lifetime of about 200~$\mu$s, producing a 2.2-MeV $\gamma$. The coincidence of these prompt and delayed events provides a distinct signal to identify reactor antineutrinos, greatly reducing backgrounds.

Reactor antineutrino candidates are selected using the criteria summarized in Table~\ref{tab:cuts}. 
A likelihood ratio (LR) is used to provide additional suppression of accidental coincidences.  Similar to the LR method used in the analysis of the SNO+ water phase~\cite{snoplus_water_antinu}, the probability density functions (PDFs) for reactor-$\overline{\nu}$ IBDs come from MC simulations, and the PDFs for accidentals are constructed from a data sample of randomly-paired events that pass the prompt and delayed event criteria. The LR is then calculated from the products of the 2D PDF of $\Delta t$ versus $\Delta r$ and the 1D PDF of the delayed event $E$. In addition, Bayesian priors are applied to account for the rates of reactor-$\overline{\nu}$ IBDs and accidentals. A cut of -3.5 on the LR was determined by maximizing a ratio of signal (S) and background (B) counts: S/$\sqrt{\text{S}+\text{B}}$. 

A suite of cuts designed to remove instrumental backgrounds is applied to all events. All data within the 20~seconds following an identified muon or high-energy event are removed to avoid cosmogenically-induced backgrounds. Additionally, any potential background from fast neutrons produced by muons in the external water is mitigated by removing events within 10~$\mu$s of an event with 3 or more hit outward-looking PMTs. Event pairs that pass the $^{214}$BiPo selection criteria in Table~\ref{tab:cuts} or have additional coincident events, are also removed. 

\begin{figure}[tbp]
    \caption{\label{fig:position_plots} Positions of selected prompt and delayed events.}
    \includegraphics[width=\columnwidth]{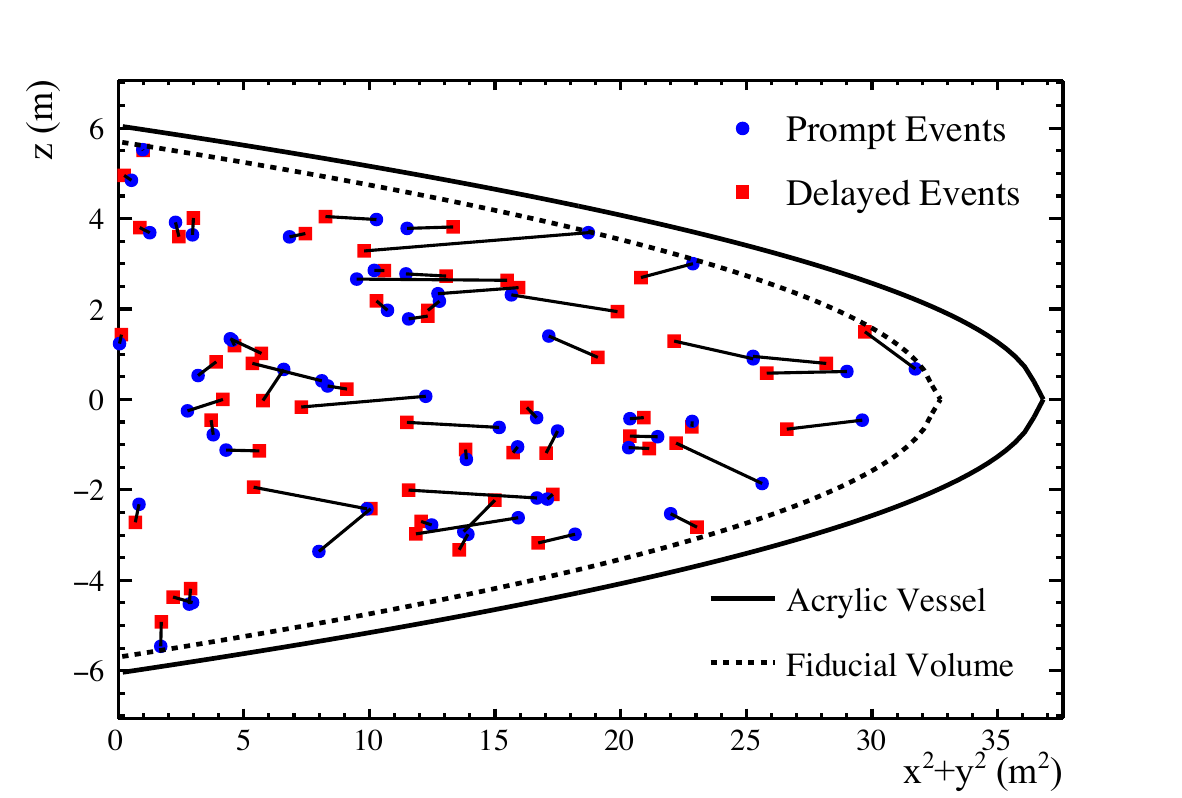}
\end{figure}

After applying all selection criteria, 59 coincidence pairs are observed in the dataset.  
The distributions of $\Delta t$, $\Delta r$, and delayed reconstructed energy of these coincidences are in good agreement with expectations for neutron captures, which were obtained by simulations of reactor-$\overline{\nu}$ IBDs~\cite{supp}.  Figure~\ref{fig:position_plots} shows the position distribution of the selected pairs, which are found to be uniform throughout the detector, as expected. 

\textit{Reactor antineutrinos}.  
Over 99\% of the expected reactor antineutrino flux at SNO+ comes from reactors in North America, around 60\% of which comes from Ontario's three Canadian Deuterium Uranium (CANDU) reactor complexes at baselines of 240~km, 340~km, and 350~km. 
Neutrino oscillations across these distances result in multiple dips in the detected prompt energy spectrum, which are well preserved given the large contribution of the CANDU reactors to the total flux. 

The flux is modeled using the thermal power outputs of reactor cores as a function of time, the average fractions of the four dominant fissile isotopes, the average energy released per fission~\cite{ma_antinu_per_mev}, and the emitted antineutrino energy spectra per isotope. 
Monthly-averaged thermal powers from yearly IAEA reports~\cite{international2022iaea, international2023iaea} are used for all reactors except the CANDU reactors for which we use the hourly electrical output provided by IESO's Generator Output and Capability reports~\cite{IESO}. 
These two reports show an average difference of $(+0.2\pm0.1)$\% in reactor power over a period of a year.  
The fission fractions of ${}^{235}$U, ${}^{238}$U, ${}^{239}$Pu and ${}^{241}$Pu depend on reactor type and also vary over time, due to the depletion of isotopes and refueling cycles. Average values of (0.568, 0.078, 0.297, 0.057)~\cite{kamland_fission_fractions} are used for the large number of pressurized water reactors (PWR) and boiling water reactors (BWR). The CANDU reactors use the pressurized heavy water reactor (PHWR) fission fractions (0.52, 0.05, 0.42, 0.01)~\cite{AECL:2013}, which are stabilized by continuous refueling and are averaged over multiple cores in each complex. 
Finally, the antineutrino energy spectrum emitted for each isotope is given by the Huber-Mueller model~\cite{huber, mueller}. The flux of the isotope model is known to be biased and is corrected by scaling by a factor of 0.945$\pm$0.007 to match the global average of reactor flux measurements~\cite{daya_bay_flux}.

Several uncertainties on the flux are carried over from the detailed studies made by the Daya Bay experiment~\cite{daya_bay_flux}. Per core, they include a 2.4\% uncertainty from the isotopic emission spectra, 0.6\% from fission fraction uncertainties, 0.5\% from power output, and several other subdominant contributions. Studies carried out by the SNO+ collaboration produced consistent results. 
The uncertainties on the shape of the predicted energy spectrum~\cite{daya_bay_flux} are expected to have a smaller impact than the detector-related uncertainties, and to be negligible.  

The number of target protons available to produce IBDs is defined by the fiducial volume.  An uncertainty on the size of the volume is determined from the position reconstruction and an uncertainty on the number of protons is inherited from the proton density in the liquid scintillator cocktail. The former is estimated to result in less than 1\% uncertainty on the volume and the latter is computed from the scintillator mass density and the hydrogen mass fractions of all included compounds. The uncertainty on mass density from measurements and temperature variations was found to be less than 0.5\%. Hydrogen mass fractions of the various carbon chains are very similar, and deviations many times larger than the manufacturer's limits would be required to produce even a modest impact on the proton density. However, Daya Bay assigned a 0.92\% uncertainty to their proton number, primarily based on deviations of these mass fractions from combustion measurements~\cite{daya_bay_flux}. In the absence of a direct independent measurement for the SNO+ cocktail, we conservatively adopt the same uncertainty.

The coincidence selection efficiency is obtained from MC simulations of IBDs occurring in the full scintillator volume, and is found to be 70\%.  Within the fiducial radius of 5.7~m, the efficiency is 83\%.  
With subpercent biases resulting from position and energy reconstruction uncertainties, all aforementioned uncertainties total to produce a 3\% uncertainty on the normalization of the reactor flux.

\textit{Neutrino oscillations}.
More than 95\% of the incoming \nuebar flux is expected to travel entirely through the North American continental crust, which has a relatively constant density~\cite{chulick2002seismic, ncei.info@noaa.gov_2012}. Therefore, the electron antineutrino survival probability is calculated using a full three-flavor mixing framework, accounting for the effect of constant matter density~\cite{mattOsc}:
\begin{align*}
    P_{ee} = 1 - 4 \sum_{n>m} (X_n)_{ee}(X_m)_{ee} \text{sin}^2\left(\left(\mathcal{E}_n - \mathcal{E}_m\right) \frac{L}{4E}\right),
\end{align*}
where $L$ and $E$ are the baseline and antineutrino energy, respectively. All other quantities are defined in Ref.~\cite{mattOsc} and depend on $\theta_{12}$, $\theta_{13}$, $\Delta m_{21}^2$, $\Delta m_{31}^2$, $E$ and the crust's electron density (8.13$\times$10$^{23}$ cm$^{-3}$).  
The matter effect induces a change in the flux of $\mathcal{O}(1\%)$ or less, and similarly, an $\mathcal{O}(1\%)$ change is induced in the effective $\widehat{\Delta m_{21}^2}$. 
For $\sin^{2}\theta_{12} = 0.307$ and $\Delta m^{2}_{21} = 7.53 \times 10^{-5}$ eV$^2$~\cite{ParticleDataGroup:2020ssz}, approximately 100 reactor-$\overline{\nu}$ IBD interactions are expected within the AV per year.

\textit{Geoneutrinos}.  
Antineutrinos up to about 3.3~MeV are produced by $\beta$ decays within the Earth. Only decays from the $^{238}$U and $^{232}$Th chains produce \nuebars with energy greater than the IBD threshold. Methods based on Ref.~\cite{Wipperfurth:2019idn} are used to predict the geoneutrino flux at SNO+ (Sudbury, Ontario) in Terrestrial Neutrino Units (TNU), where 1~TNU corresponds to 1 IBD interaction per year per 10$^{32}$ free protons. Assuming a constant survival probability of $<P_{ee}> = 0.55$ 
and a radiogenic heat of 20~TW, our model predicts geo-$\overline{\nu}$ IBD rates of 36.3$\pm$8.7 TNU and 9.7$\pm$2.3 TNU from the $^{238}$U and $^{232}$Th chains, respectively. This gives an expected rate of 27 geo-$\overline{\nu}$ IBDs within the AV per year. 
Recognizing uncertainty in the range of possible Earth radiogenic heat values and in the local geology, the total geo-$\overline{\nu}$ IBD rate is fit without a direct constraint.  
As in geoneutrino studies by KamLAND and Borexino, knowledge of Earth’s Th and U chondritic abundances~\cite{Wipperfurth:2018} motivates applying a constraint on the ratio of U/Th geo-$\overline{\nu}$ IBD event rates when fitting the data.  In this analysis, we use the predicted fluxes from our model to derive a constraint on the U/Th ratio of 3.7 $\pm$ 1.3, where the uncertainty comes from combining each of the U and Th flux prediction uncertainties.

\textit{\alphan backgrounds}.  
The dominant background in this analysis is from \alphan interactions with the natural $^{13}$C present in the scintillator. These interactions produce $^{16}$O and a neutron, which thermalizes and captures, mimicking the IBD delayed signal. The prompt event can proceed through three channels, each producing a distinct energy peak. The neutron can elastically scatter protons, producing a signal in the apparent energy range of roughly 0.5~MeV to 3.5~MeV, or it can inelastically scatter off a $^{12}$C, which emits a 4.4-MeV $\gamma$. Alternatively, the $^{16}$O can be produced in an excited state, de-exciting by producing either a 6-MeV $\gamma$ or an electron-positron pair that deposits about 6~MeV. 

The dominant source of $\alpha$ decays in the detector is $^{210}$Po. The rate of these decays is measured over time with a fit of the $\alpha$'s energy peak, which quenches to be around 0.4~MeV.  
The \alphan background from $^{210}$Po implanted on the surface of the AV is reduced to a negligible level by the fiducial volume selection.  
The average rate within the fiducial volume over this dataset is 38~Hz, nearly a factor of 5 reduction in specific $^{210}$Po activity compared to the initial oscillation measurement, as described in Ref.~\cite{snoplus_partial_antinu}.  

The probability of the $^{210}$Po $\alpha$ undergoing an \alphan interaction is calculated by integrating the energy-dependent interaction cross section up to the $^{210}$Po $\alpha$ energy, and then multiplying by the $^{13}$C number density, in the same way as Ref.~\cite{snoplus_partial_antinu}. This, combined with the average $^{210}$Po rate, gives an average expected \alphan rate of 0.21~events/day.  

Disagreements between total cross section measurements and particularly large uncertainties in branching ratios (ground state vs. excited state) warrant a conservative approach to \alphan uncertainties~\cite{Page2025}. The uncertainties from Ref.~\cite{snoplus_partial_antinu} are assigned to the normalizations of the three \alphan interaction channels: for proton elastic scatters and the $^{12}$C inelastic scatter, 30\% uncertainties are assigned, and for the excited $^{16}$O channel, a 100\% uncertainty is assigned.

\textit{$^{214}$BiPo-like background}. 
Immediately after scintillator filling operations, the data showed higher rates of radioactive backgrounds primarily due to ingress of radon, resulting in most of these backgrounds decaying away with the $^{222}$Rn half-life of 3.8~days.  
During these high-background periods, an excess of coincidences was observed with delayed event energies just below the 2.2-MeV region of interest. 
The $\Delta t$ and $\Delta r$ distributions of these coincidences were consistent with a correlated decay of time and distance similar to those of IBD and $^{214}$BiPo coincident events, and the position distribution was uniform.  

The prompt energy distribution of these events is in good agreement with the $^{214}$Bi $\beta$ decay spectrum (Q = 3.3~MeV) that precedes a $^{214}$Po $\alpha$ decay, which has an energy of 7.8~MeV, but quenches down to about 0.8~MeV in visible energy in the scintillator.  
The half-life of the $^{214}$Po decay is 164~$\mu$s, which is close to the approximately 210-$\mu$s neutron capture time for IBDs.  
Rare $\alpha + \gamma$ decays of $^{214}$Po are too low in energy to create the excess.
A likely explanation is that these tails arise from alpha-proton elastic scattering interactions, in which more scintillation light is produced than by the $\alpha$'s alone due to the lesser quenching of protons. Data show that $\alpha$ decays of $^{215}$Po exhibit a similar tail extending to higher energies.

\begin{figure}[tbp]
    \centering
    \caption{\label{fig:sideband_background}Fit of extended delayed energy distribution.  The model includes a data-driven kernel density estimate of the $^{214}$Po spectrum based on the observed spectrum of $^{215}$Po. $^{214}$Po backgrounds, which dominate the lower half of the spectrum, were increased in this plot relative to the analyzed data via the inclusion of data from the period immediately after scintillator fill operations, when radon levels in the detector were elevated.}
   \includegraphics[width=\columnwidth]{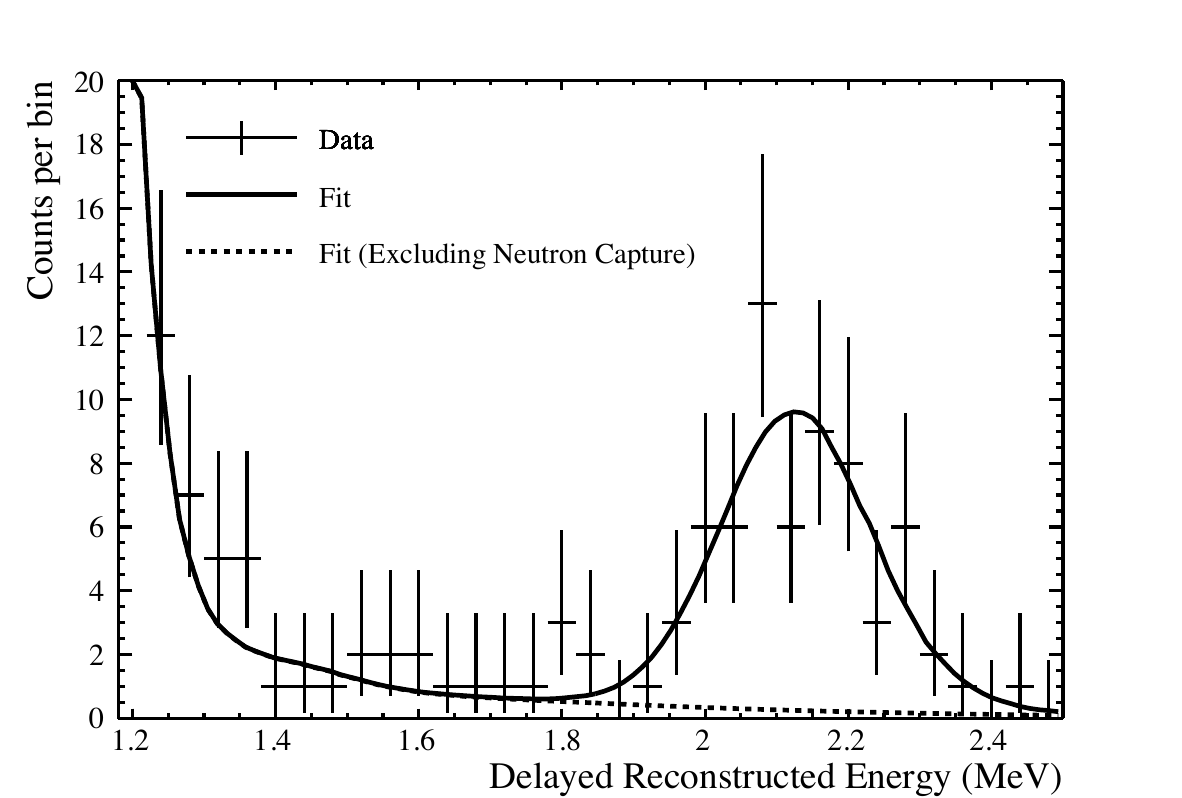}
\end{figure}

For this analysis, a data-driven model of the $^{214}$Po energy spectrum was created using a kernel density estimation of the $^{215}$Po energy distribution.  
Figure~\ref{fig:sideband_background} shows that the model fits well to the delayed energy spectrum, which is dominated by $^{214}$Po in the lower half. The corresponding prompt energy is modeled as $^{214}$Bi.
The LR can help suppress this type of background due to the differences in the delayed energy distributions between $^{214}$Po and 2.2-MeV neutron capture $\gamma$'s.  
Across the dataset used in the present analysis, the model predicts that 1.1$\pm$1.1 $^{214}$BiPo coincidence events will pass the IBD selection criteria, which has a small effect on sensitivity estimates for $\Delta m^2_{21}$. This was also confirmed for different spectral shapes\cite{Zummo:2024fbp}.

\textit{Other backgrounds}.  
The rate of accidental coincidences is calculated using the measured rates of events passing the selection criteria for prompt and delayed event candidates.  After applying all selection criteria, including the LR cut, the calculation gives an expectation of 0.3 coincidences in the entire 134.4-day dataset.  Since it is based directly on data, the prediction has a negligible uncertainty.  

Atmospheric neutrinos can undergo neutral- and charged-current interactions with the hydrogen and carbon present in the scintillator, which can produce neutrons and mimic the IBD signal.
A study of simulated atmospheric neutrino interactions at SNO+ was performed using the GENIE Monte Carlo generator 
\cite{andreopoulos2010genie} and found a negligible contribution relative to the expected IBD signal. 

Cosmogenic muon interactions in the detector can produce neutrons and sources of ($\beta-n$) decays. These backgrounds are reduced to a negligible level by the muon veto cuts.

\begin{figure}[tbp]
    \centering
    \caption{\label{fig:Eprompt}Energy distribution of prompt events and best-fit (unconstrained) predictions.}
    \includegraphics[width=\columnwidth]{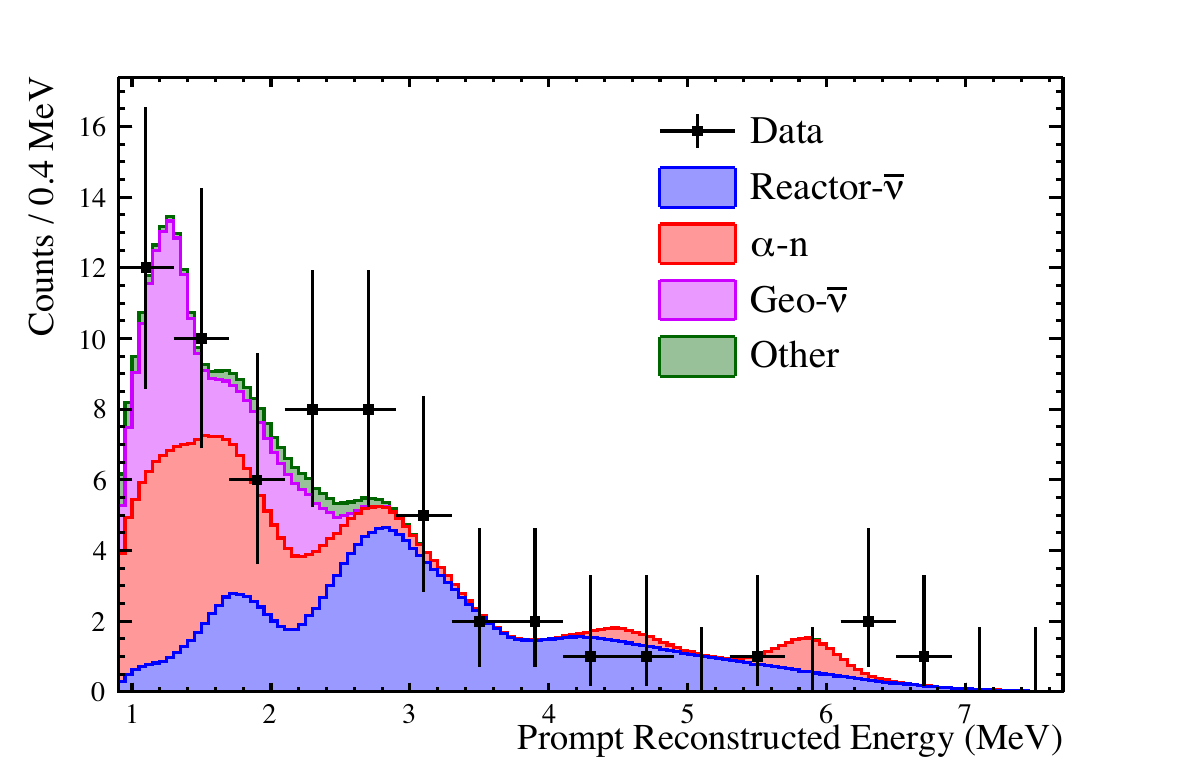}
\end{figure}

\textit{Spectral Analysis and Results}.  
The expected numbers of signal and background events are listed in Table~\ref{tab:sum_sig_bkg} and the corresponding prompt energy spectra are shown in Fig.~\ref{fig:Eprompt}. 
The total expected number of selected coincidences is 54.7, in reasonable agreement with the 59 coincidences observed in the data.  
Two independent analyses were performed to cross-check this event selection and the following fits.

\begin{table}[tbp]\centering
    \caption{\label{tab:sum_sig_bkg}Expected and fitted numbers of signal and background events, and the total number of observed events in 134.4 days of data. Expectations show only systematic uncertainties and the geo-$\overline{\nu}$ IBD rate is unconstrained. The fits are with unconstrained or constrained oscillation parameters.}
\begin{tabular}{lccc}
\toprule
\toprule
                    &  Expectation   & Fit (Uncon.)  & Fit (Con.)\\
\midrule
Reactor-$\overline{\nu}$ IBD         & 27.9$\pm$0.8  & 25.1$^{+6.4}_{-2.1}$ & 27.5$\pm$0.9 \\
\alphan             & 18.2$\pm$5.2  & 17.2$^{+4.5}_{-4.4}$ & 17.2$^{+4.5}_{-4.4}$\\
Geo-$\overline{\nu}$ IBD             & 7.2           & 12.0$^{+7.4}_{-6.8}$ & 11.1$^{+7.1}_{-6.6}$ \\ 
$^{214}$BiPo-like  & 1.1$\pm$1.1   & 1.2$\pm$1.1          & 1.2$\pm$1.1\\
Accidental          & 0.3$\pm$0.0   & 0.3$\pm$0.0          & 0.3$\pm$0.0\\
\midrule
Total               & 54.7          & 55.8         & 57.3\\
\midrule 
Observed            & 59            & 59                   & 59\\
\bottomrule
\bottomrule
\end{tabular}
\end{table}

An extended $\log$ likelihood fit to unbinned data was performed on the prompt energy distribution to find the best-fit values of $\Delta m^{2}_{21}$ and $\sin^2\theta_{12}$. 
The data and fitted energy spectra are shown in Fig.~\ref{fig:Eprompt}. 
Nuisance parameters were constrained with Gaussian penalty terms added to the likelihood function. These parameters include the normalizations on reactor-$\overline{\nu}$ IBDs, \alphan channels, and the geoneutrino U/Th ratio, as well as energy systematics.

The uncertainties on the prompt energy are dominated by reconstruction uncertainties~\cite{Zummo:2024fbp, Page2025}. The energy systematics are modeled as uncertainties in the energy scale -- both linear and non-linear -- and in the energy resolution. Two independent factors parameterize the linear energy scale uncertainties of $\beta$'s/$\gamma$'s (1.8\%) and protons (3\%) to account for their different scintillation response and prediction uncertainties. A nonlinear scaling following Birks' law is applied as a variation in Birks' constant $k_B$ (5.4\%). Finally, energy resolution uncertainty is modeled as a Gaussian smearing with a standard deviation of 4.4\%$\times\sqrt{E}$.

\begin{figure}[tbp]
    \centering
    \caption{\label{fig:fit_results}The $\log$ likelihood of the prompt energy spectrum as a function of $\Delta m^2_{21}$ vs. $\sin^2\theta_{12}$ with 1$\sigma$ and 2$\sigma$ contours drawn. The $\log$ likelihood is also shown separately for each variable, fixing the other to the best-fit value. The slight asymmetry about 0.5 is expected and gives a second best-fit point for $\sin^2\theta_{12}$ near the global result 0.307~\cite{ParticleDataGroup:2020ssz}.}
   \includegraphics[width=\columnwidth]{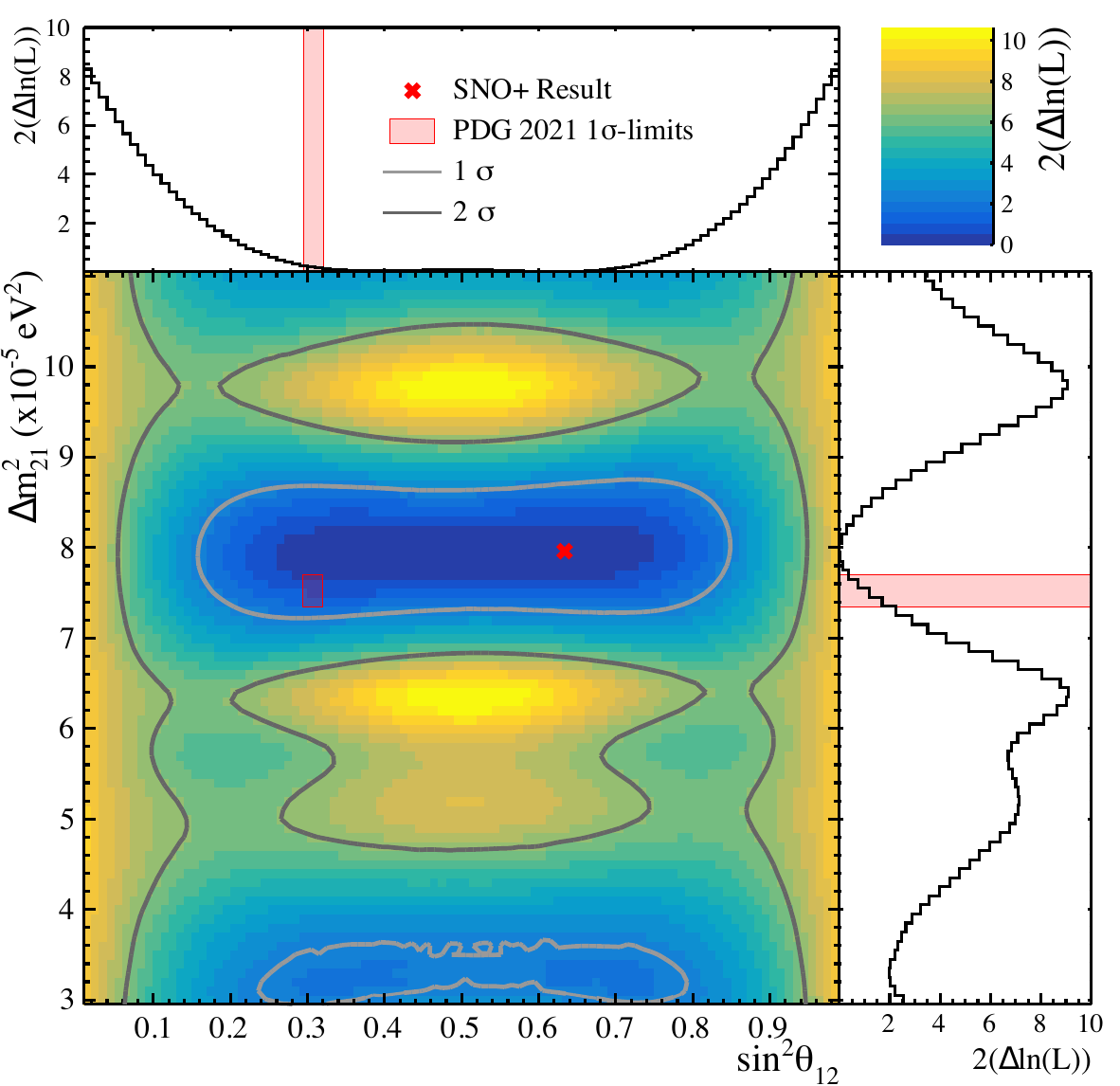}
\end{figure}

A 2D visualization of the measurement significance is shown in Fig.~\ref{fig:fit_results} as a grid scan of the $\log$ likelihood over $\Delta m^{2}_{21}$ and $\sin^2\theta_{12}$, profiling over all other parameters. Performing a complete fit to account for correlations between all parameters, the best-fit values for the unconstrained oscillation parameters are $\Delta m^{2}_{21}$ = \dmResult eV$^2$ and $\sin^2\theta_{12}$ = \sstResRight.  
A nearly equally significant result for $\sin^2\theta_{12}$ occurs at the mirror value across 0.5. 
Table~\ref{tab:sum_sig_bkg} shows the associated best-fit numbers of the signals and backgrounds.

Combining the previous measurement from SNO+~\cite{snoplus_partial_antinu} with the present result produces minimal change; therefore, it has not been included.  
The result for $\Delta m^2_{21}$ agrees with the KamLAND result, (7.53 $\pm$ 0.18) $\times$ 10$^{-5}$~eV$^2$~\cite{kamland_on_off}, within 1$\sigma$. 
It shows a slightly worse agreement of 1.3$\sigma$ with the combined solar result from Super-K, (6.10$^{+0.95}_{-0.81}$) $\times$ 10$^{-5}$~eV$^2$~\cite{Super-Kamiokande:2023jbt}. 

The fit is repeated assuming Gaussian constraints of $\Delta m^{2}_{21} = (7.53 \pm 0.18) \times 10^{-5}$ eV$^2$~\cite{kamland_on_off} and $\sin^2\theta_{12} = 0.307 \pm 0.013$~\cite{superk_solar_old}. 
The fitted number of geo-$\overline{\nu}$ IBDs is 11.1$^{+7.1}_{-6.6}$, which corresponds to a geo-$\overline{\nu}$ IBD rate of \geoResult TNU. 
The measured geoneutrino rate has a relatively large uncertainty, making it consistent with the range of existing Earth models allowed by the two previous measurements. 
The uncertainty in the geo-$\overline{\nu}$ IBD rate is dominated by a negative correlation with the proton scattering component of \alphan and is not significantly improved by constraining the oscillation parameters. 
The resulting best-fit oscillation parameters are $\Delta m^{2}_{21}$ = $7.58^{+0.18}_{-0.17}$ $\times$ 10$^{-5}$~eV$^2$ and $\sin^2\theta_{12} = 0.308 \pm 0.013$. 
 
The tension between solar and reactor results for $\Delta m^{2}_{21}$, after combining SNO+ and KamLAND, is slightly greater than 1.5$\sigma$.  
Table~\ref{tab:fit_results} summarizes the results of the two different fits. All nuisance parameters are found to be in good agreement with expectations in both cases.  
Given that $\sin^2\theta_{12}$ is largely determined by the flux and $\Delta m^2_{21}$ by the spectrum, the two parameters are minimally correlated, leaving little to gain by constraining one and fitting the other. 

\begin{table}[tbp]
    \centering
    \caption{\label{tab:fit_results}Best-fit values for oscillation parameters and geo-$\overline{\nu}$ IBD rate. Results are reported with no constraints on oscillation parameters $\Delta m^2_{21}$ and $\sin^2\theta_{12}$, and with Gaussian constraints according to their current best measurements (see text).}
\begin{tabular}{lccc}
\toprule
\toprule
            & Fit (Uncon.)    & Fit (Con.)\\
\midrule
$\Delta m^2_{21}$ ($\times 10^{-5} \text{eV}^2$)  & $7.96^{+0.48}_{-0.42}$  & $7.58^{+0.18}_{-0.17}$\\
$\sin^2\theta_{12}$     & \sstResRight  & $0.308\pm0.013$\\
Geo-$\overline{\nu}$ IBD rate (TNU)      & $79^{+49}_{-44}$  & \geoResult\\
\bottomrule
\bottomrule
\end{tabular}
\end{table}

\textit{Outlook}. 
Since the acquisition of the data used in this analysis, a secondary fluor (bis-MSB) was added to the SNO+ scintillator, increasing the detected light by more than 50\%.  The SNO+ search for neutrinoless double beta decay is scheduled to begin with the loading of tellurium near the end of 2025, and will allow the continued analysis of antineutrinos.

With its unique pattern of long-baseline reactor antineutrino oscillations, SNO+ will continue to update its independent measurement of $\Delta m^{2}_{21}$, to compare and combine with those from dedicated experiments, such as KamLAND and JUNO~\cite{juno2022}. 

The SNO+ collaboration has identified a $^{214}$BiPo-like background component that is likely due to alpha-proton scattering and that has the potential to affect many experiments measuring IBDs.  As more data are collected, this and the dominant \alphan background will be better characterized. The use of pulse shape discrimination to suppress the \alphan is in preparation and is predicted to significantly improve the experiment's sensitivity to the geo-$\overline{\nu}$ IBD rate. 

With a larger flux of geoneutrinos expected from the thick North American plate, SNO+ measurements are highly complementary to observations at shallower locations, improving the separation of the crust and mantle fluxes most relevant for Earth modeling.

\textit{Summary}.
With 286 tonne-years of data, the SNO+ collaboration reports the second-most precise measurement of $\Delta m^{2}_{21}$, providing an independent evaluation of long-baseline reactor antineutrino oscillations. We find $\Delta m^{2}_{21}$ = \dmResult eV$^2$  
while also measuring $\sin^2\theta_{12}$ = \sstResRight. 
Combining with previous measurements from long-baseline reactor antineutrinos and solar neutrinos yields $\Delta m^{2}_{21}$ = $7.58^{+0.18}_{-0.17}$ and a measurement of the geo-$\overline{\nu}$ IBD rate at SNO+: \geoResult TNU.

\begin{acknowledgments}
Capital funds for SNO\raisebox{0.5ex}{\tiny\textbf{+}} were provided by the Canada Foundation for Innovation and matching partners: 
Ontario Ministry of Research, Innovation and Science, 
Alberta Science and Research Investments Program, 
Queen’s University at Kingston, and 
the Federal Economic Development Agency for Northern Ontario. 
This research was supported by 
{\it Canada: }
the Natural Sciences and Engineering Research Council of Canada, 
the Canadian Institute for Advanced Research, 
the Ontario Early Researcher Awards, 
the Arthur B. McDonald Canadian Astroparticle Physics Research Institute; 
{\it U.S.: }
the Department of Energy (DOE) Office of Nuclear Physics, 
the National Science Foundation and the DOE National Nuclear Security
Administration through the Nuclear Science and Security Consortium; 
{\it UK: }
the Science and Technology Facilities Council and the Royal Society; 
{\it Portugal: } 
Funda\c{c}\~{a}o para a Ci\^{e}ncia e a Tecnologia (FCT-Portugal); 
{\it Germany: }
the Deutsche Forschungsgemeinschaft; 
{\it Mexico: }
DGAPA-UNAM and Consejo Nacional de Ciencia y Tecnolog\'{i}a; 
{\it China: }
the Discipline Construction Fund of Shandong University.  
We also thank SNOLAB and SNO\raisebox{0.5ex}{\tiny\textbf{+}} technical
staff; the Digital Research Alliance of Canada; the
GridPP Collaboration and support from Rutherford
Appleton Laboratory; and the Savio computational cluster
at the University of California, Berkeley. Additional long-term
storage was provided by the Fermilab Scientific Computing
Division.

\end{acknowledgments}

\bibliographystyle{apsrev}
\bibliography{References}

\end{document}



{\centering{}
\LARGE{SNO+ Experiment} \\ 
\Large{Supplemental Material} \\
}

\bigskip
\bigskip
\normalsize
This document provides supplemental information relevant to the measurement of reactor antineutrinos oscillation in ``Measurement of reactor antineutrino oscillation at SNO+''.





\begin{table}[ht!]
    \centering
    \caption{Information about the 59 coincidence pairs selected.}
    \label{tab:event_info}
    \begin{tabular}{crrr|crrr|rrr} \hline \hline 
         \multicolumn{4}{c|}{Prompt} & \multicolumn{4}{c|}{Delayed} & \\ \hline 
         $E$ (MeV) & $x$ (m) & $y$ (m) & $z$ (m) & $E$ (MeV) & $x$ (m) & $y$ (m) & $z$ (m) & $\Delta t$ ($\mu$s) & $\Delta r$ (cm) & Date \\ \hline 

1.81 & -2.81 & -3.10 & -0.70 & 1.93 & -2.97 & -2.86 & -1.19 & 73.92 & 56.30 & 20/05/2022\\
3.23 & 2.40 & -2.83 & -2.92 & 2.09 & 2.53 & -2.93 & -2.23 & 59.34 & 71.60 & 23/05/2022\\
1.16 & -2.10 & 3.35 & 2.31 & 2.13 & -2.38 & 3.77 & 1.94 & 330.78 & 62.42 & 26/05/2022\\
1.66 & -1.02 & -0.81 & -5.46 & 2.30 & -1.21 & -0.50 & -4.92 & 294.16 & 65.10 & 27/05/2022\\
2.58 & 0.61 & -1.68 & 0.53 & 2.17 & 0.67 & -1.86 & 0.83 & 132.10 & 35.95 & 27/05/2022\\
1.89 & 1.34 & -1.09 & -4.49 & 2.11 & 1.17 & -0.91 & -4.37 & 23.72 & 27.47 & 28/05/2022\\
1.50 & 1.04 & -4.92 & 0.90 & 1.98 & 1.02 & -4.59 & 1.29 & 133.82 & 51.06 & 28/05/2022\\
6.38 & 3.37 & 4.51 & 0.68 & 2.06 & 2.69 & 4.74 & 1.50 & 235.42 & 108.53 & 30/05/2022\\
1.02 & -2.57 & -2.22 & 1.78 & 2.28 & -2.56 & -2.41 & 1.84 & 83.44 & 19.45 & 31/05/2022\\
1.77 & -1.92 & -0.78 & -1.12 & 2.03 & -2.13 & -1.05 & -1.14 & 300.56 & 33.82 & 3/06/2022\\
2.29 & 2.18 & 3.45 & -0.40 & 2.11 & 2.06 & 3.46 & -0.18 & 15.88 & 25.45 & 9/06/2022\\
1.03 & -0.23 & 0.04 & 1.23 & 2.21 & -0.36 & 0.00 & 1.43 & 39.02 & 24.28 & 9/06/2022\\
2.68 & 1.95 & -2.76 & 2.78 & 2.18 & 2.25 & -2.82 & 2.73 & 74.48 & 30.80 & 13/06/2022\\
6.89 & -1.15 & -3.72 & -0.62 & 2.09 & -0.98 & -3.25 & -0.51 & 187.18 & 51.65 & 19/06/2022\\
1.14 & 1.95 & 4.21 & -0.83 & 2.22 & 1.94 & 4.07 & -0.81 & 45.3 & 13.27 & 20/06/2022\\
1.92 & 4.28 & -1.43 & -0.43 & 2.04 & 4.35 & -1.41 & -0.40 & 479.34 & 7.49 & 21/06/2022\\
0.96 & 2.57 & 0.05 & 0.66 & 2.24 & 2.39 & -0.25 & -0.03 & 347.6 & 77.45 & 28/06/2022\\
1.88 & -0.38 & 5.43 & -0.46 & 2.07 & -0.23 & 5.15 & -0.66 & 13.62 & 37.13 & 30/06/2022\\
2.35 & -2.92 & -0.99 & 2.66 & 2.05 & -3.71 & -1.31 & 2.63 & 9.72 & 85.45 & 2/07/2022\\
2.28 & -3.56 & -3.19 & 3.00 & 2.13 & -3.19 & -3.26 & 2.70 & 86.3 & 48.38 & 2/07/2022\\
5.58 & 0.78 & -1.96 & 1.34 & 2.27 & 0.97 & -2.18 & 1.02 & 18.48 & 43.41 & 3/07/2022\\
3.14 & -3.83 & -3.31 & -1.86 & 2.27 & -3.57 & -3.07 & -0.97 & 579.92 & 96.01 & 30/07/2022\\
2.22 & 2.75 & 2.51 & -1.33 & 1.95 & 2.81 & 2.44 & -1.11 & 429.2 & 23.44 & 23/08/2022\\
4.59 & 2.07 & 1.93 & -3.36 & 2.13 & 2.17 & 2.31 & -2.42 & 181.1 & 102.55 & 1/09/2022\\
1.06 & -1.11 & -3.32 & 0.07 & 2.27 & -0.82 & -2.57 & -0.17 & 90.56 & 83.50 & 2/09/2022\\
2.94 & -2.75 & 0.88 & 0.30 & 2.01 & -2.88 & 0.90 & 0.23 & 143.30 & 14.93 & 13/09/2022\\
6.12 & -3.09 & -1.72 & -2.78 & 2.07 & -2.89 & -1.93 & -2.70 & 267.00 & 29.87 & 16/09/2022\\
4.45 & -2.04 & 2.94 & 2.17 & 2.09 & -1.86 & 2.98 & 1.97 & 530.58 & 27.28 & 17/10/2022\\
1.45 & -2.83 & 1.64 & 1.97 & 2.07 & -2.90 & 1.37 & 2.18 & 212.66 & 35.09 & 27/10/2022\\
1.19 & 3.00 & -4.47 & 0.62 & 2.20 & 2.66 & -4.33 & 0.58 & 464.34 & 37.56 & 3/11/2022\\
2.90 & -1.38 & -0.61 & 3.92 & 2.07 & -1.36 & -0.75 & 3.60 & 182.32 & 34.52 & 4/11/2022\\
1.11 & 0.16 & 1.94 & -0.78 & 2.25 & 0.26 & 1.91 & -0.46 & 411.46 & 34.01 & 6/11/2022\\
0.95 & -2.07 & -3.80 & 3.69 & 2.17 & -1.90 & -2.49 & 3.28 & 279.48 & 138.17 & 8/11/2022\\
1.27 & 3.00 & 2.85 & 1.40 & 1.98 & 2.93 & 3.24 & 0.93 & 23.20 & 61.39 & 9/11/2022\\
1.56 & 0.05 & 0.73 & 4.85 & 2.05 & 0.16 & 0.46 & 4.95 & 300.60 & 30.62 & 11/11/2022\\
1.39 & 1.64 & -0.29 & -0.25 & 2.19 & 1.98 & -0.48 & 0.00 & 31.90 & 47.12 & 26/11/2022\\
2.90 & -0.56 & 4.75 & -0.49 & 2.14 & -0.51 & 4.75 & -0.60 & 13.74 & 12.53 & 27/11/2022\\
2.04 & 0.30 & 3.18 & 2.85 & 1.96 & 0.02 & 3.25 & 2.85 & 210.38 & 29.45 & 9/12/2022\\
2.42 & -1.97 & 2.45 & -2.42 & 2.29 & -1.72 & 1.57 & -1.94 & 156.88 & 104.04 & 11/12/2022\\
1.66 & -4.19 & 2.11 & -2.53 & 2.35 & -4.33 & 2.07 & -2.83 & 952.90 & 33.14 &  11/12/2022\\
1.25 & 0.05 & -1.12 & 3.69 & 2.13 & 0.17 & -0.91 & 3.80 & 82.82 & 26.61 & 25/12/2022\\
2.14 & -0.43 & 0.80 & -2.32 & 2.02 & 0.03 & 0.83 & -2.72 & 284.56 & 61.05 & 25/12/2022\\
2.84 & 3.26 & 0.93 & 3.78 & 2.33 & 3.59 & 0.69 & 3.81 & 511.54 & 40.21 & 29/12/2022\\
2.88 & -3.54 & -0.48 & 2.34 & 2.20 & -3.98 & -0.30 & 2.47 & 84.76 & 49.63 & 4/01/2023\\
3.47 & 3.85 & 1.04 & -2.62 & 2.15 & 3.31 & 0.96 & -2.97 & 7.00 & 65.82 & 4/01/2023\\
3.98 & 0.25 & 0.96 & 5.52 & 2.27 & 0.43 & 0.90 & 5.51 & 9.96 & 19.04 & 9/01/2023\\
2.10 & 0.55 & -1.63 & 3.64 & 2.08 & 0.90 & -1.48 & 4.01 & 5.44 & 53.56 & 9/01/2023\\
2.86 & -1.14 & -3.97 & -2.20 & 2.20 & -1.22 & -3.98 & -2.10 & 137.68 & 13.29 & 22/01/2023\\
3.58 & 3.73 & -3.37 & 0.95 & 2.17 & 3.94 & -3.56 & 0.80 & 5.42 & 32.20 & 27/01/2023\\
4.01 & 2.99 & -3.04 & -2.98 & 2.16 & 2.59 & -3.16 & -3.17 & 22.44 & 45.28 & 27/01/2023\\
1.61 & 1.57 & -4.23 & -1.07 & 2.22 & 1.69 & -4.27 & -1.08 & 371.46 & 13.48 & 28/01/2023\\
1.58 & -1.22 & 2.96 & 3.98 & 1.99 & -1.27 & 2.58 & 4.04 & 36.40 & 39.56 & 29/01/2023\\
1.40 & -1.96 & -3.58 & -2.18 & 2.44 & -1.51 & -3.05 & -2.00 & 52.44 & 72.07 & 3/02/2023\\
1.01 & 1.40 & 0.93 & -4.53 & 2.09 & 1.34 & 1.04 & -4.19 & 199.10 & 36.07 & 6/02/2023\\
2.91 & -0.92 & -1.92 & 1.31 & 2.01 & -0.90 & -1.95 & 1.19 & 229.16 & 12.36 & 11/02/2023\\
2.22 & 3.54 & 1.16 & -2.98 & 2.03 & 3.46 & 1.25 & -3.33 & 445.90 & 36.39 & 14/02/2023\\
2.93 & -1.58 & 2.37 & 0.41 & 2.10 & -0.97 & 2.10 & 0.80 & 212.56 & 77.33 & 16/02/2023\\
2.84 & -1.36 & 3.75 & -1.05 & 2.15 & -1.33 & 3.73 & -1.18 & 58.36 & 13.12 & 25/02/2023\\
1.68 & 1.63 & 2.04 & 3.59 & 2.10 & 1.73 & 2.11 & 3.67 & 341.78 & 14.24 & 26/02/2023\\

\hline \hline

    \end{tabular}
\end{table}
\clearpage

\section*{Event distributions}

\begin{figure}[htbp]
\caption{\label{fig:delayed_pdfs} Time between prompt and delayed events ($\Delta t$) for both data and reactor IBD simulation. The exponential fit (red) results in a neutron capture time constant of (197 $\pm$ 32)~$\mu$s. 
}
\includegraphics[width=0.7\columnwidth]{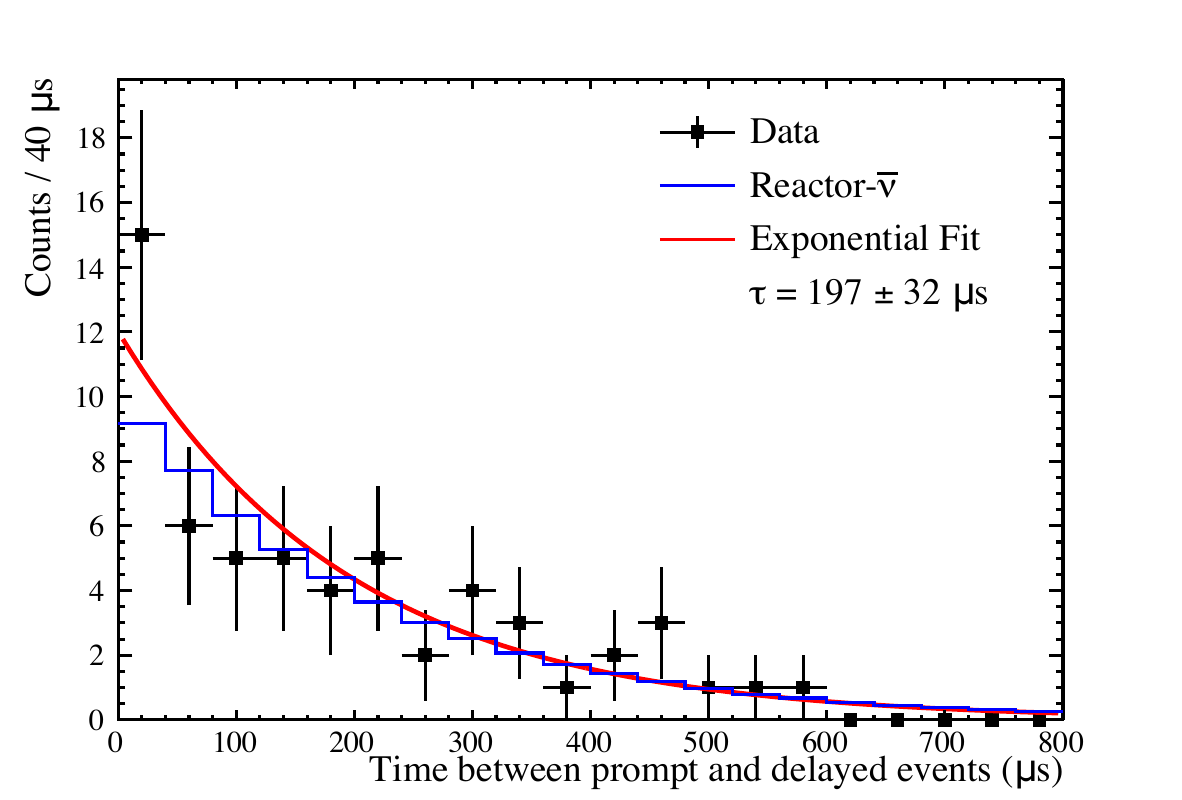}
\end{figure}

~\\

\begin{figure}[htbp]
\caption{\label{fig:delayed_pdfs} Distance between prompt and delayed events ($\Delta r$) for both data and reactor IBD simulation.}
\includegraphics[width=0.7\columnwidth]{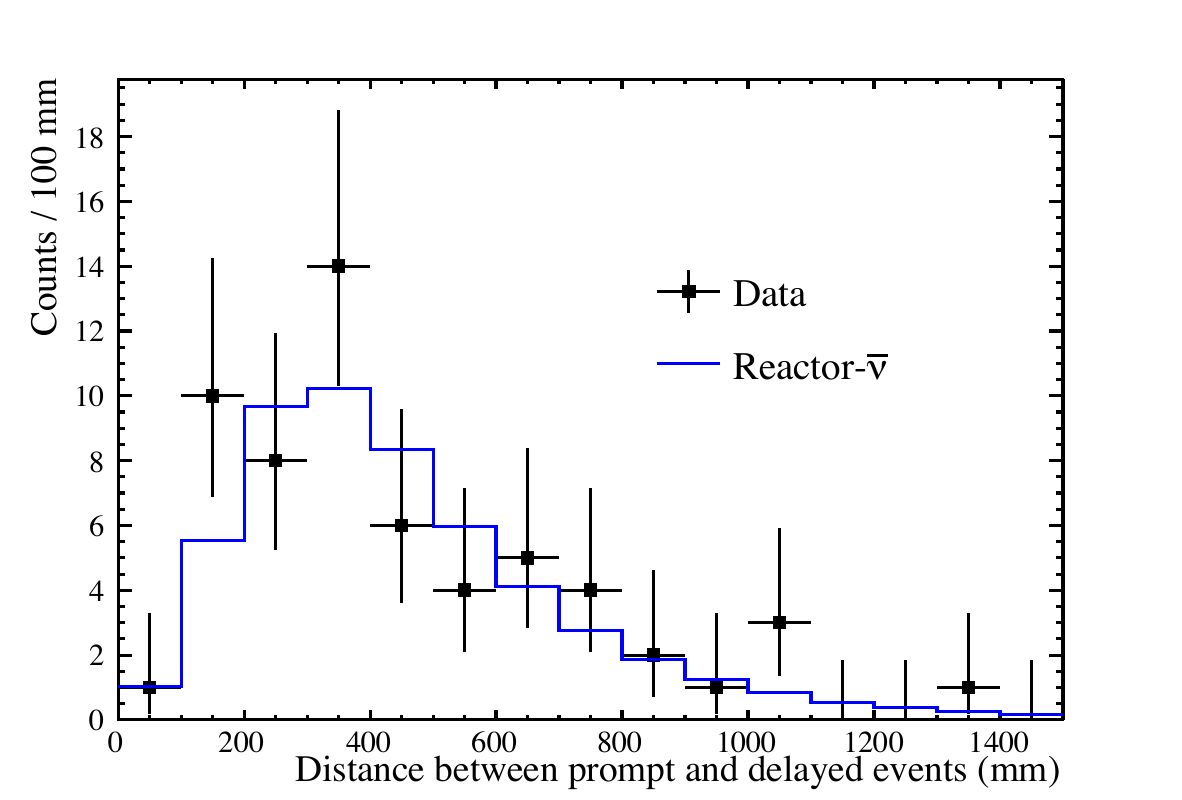}
\end{figure}

\begin{figure}[htbp]
\caption{\label{fig:delayed_pdfs} Delayed event energy ($E$) for both data and reactor IBD simulation.}
\includegraphics[width=0.7\columnwidth]{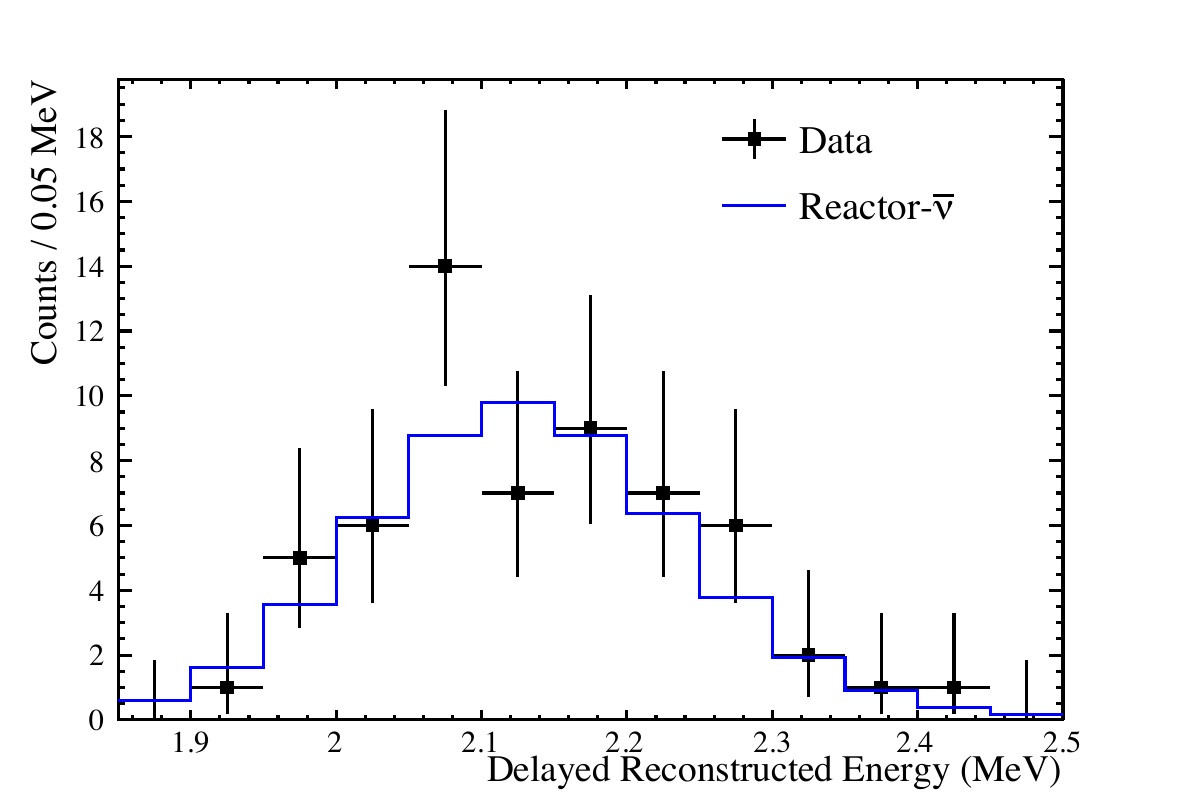}
\end{figure}